# Why liquids are fragile

R.Casalini[1,2,@] and C.M.Roland[1,*]

[1]*Naval Research Laboratory, Code 6120, Washington DC  20375-5342*
[2]*George Mason University, Fairfax VA 22030*

[@] e-mail: casalini@nrl.navy.mil, [*] e-mail: roland@nrl.navy.mil

( March 18 2005 )

Abstract: The fragilities ($T_g$-normalized temperature dependence of α-relaxation times) of 33 glass-forming liquids and polymers are compared for isobaric, $m_P$, and isochoric, $m_V$, conditions. We find that the two quantities are linearly correlated, $m_P = (37 \pm 3) + (0.84 \pm 0.05) m_V$. This result has obvious important consequences, since the ratio $m_V/m_P$ is a measure of the relative degree to which temperature and density control the dynamics, Moreover, we show that the fragility itself is a consequence of the relative interplay of temperature and density effects near $T_g$. Specifically, strong behavior reflects a substantial contribution from density (jammed dynamics), while the relaxation of fragile liquids is more thermally-activated. Drawing on the scaling law $\log(\tau) = \Im(T\upsilon^\gamma)$, a physical interpretation of this result in terms of the intermolecular potential is offered.

The glass transition remains one of the more intriguing topics in condensed matter physics, with much effort focused on understanding the progressive slowing down of the dynamics. This proceeds over more than ten decades in time, with the supercooled liquid eventually arriving in a non-equilibrium state below its glass temperature, $T_g$. Efforts to probe this feature of vitrifying liquids often employ the fragility,

$$m = \left.\frac{d\log(x)}{d(T_g/T)}\right|_{T=T_g} \quad (1)$$

as a measure of the effect of temperature on the dynamics. In eq.1 $T$ is the absolute temperature, $x$ can be the relaxation time ($\tau$) or inverse viscosity ($\eta^{-1}$), and $T_g$ is commonly defined as the temperature at which $x$ assumes some arbitrary value (e.g., $\tau = 100$ s or $\eta = 10^{12}$ Pa s). The term fragility was coined by Angell[1,2,3] to refer to the loss of the local structure (short range order) with increasing $T$ across the glass transition. For fragile liquids this structure is rapidly "broken", and large changes in $x$ with $T_g/T$ are observed. Strong liquids maintain their short range order to higher temperatures, with consequent smaller changes in $x$ (this property makes them preferable for glass-blowing).

While there are other ways to quantify the temperature dependence of a glass-former's dynamics, fragility correlates with many other properties[4,5,6,7], even those having characteristic times much shorter than the timescale for structural relaxation.[8,9,10,11,12,13,14] Fragility also serves as the basis for some theoretical interpretations of the glass transition.[15,16,17,18] In this paper, we make use of recent results, in particular data for high pressure by ourselves and other groups, to offer an alternative interpretation of fragility. In conventional isobaric measurements, the only experimental variable is temperature, and thus thermal energy and volume effects are convoluted. However, high pressure measurements in combination with the equation of state (EOS) allow characterization of a material at constant temperature and varying volume (that is, specific volume, $\upsilon$), whereby the relative effects of temperature and $\upsilon$ on the dynamics can be quantified.

Figure 1(a) shows typical behavior for the $\upsilon$–dependence of dielectric relaxation times $\tau$ measured at atmospheric and high pressure under isothermal conditions. The materials are a polychlorinated biphenyls (PCB54)[19] and propylene carbonate (PC)[20], which represent rather extreme cases in temperature and specific volume effects. Fig 1(a) shows clearly that neither $T$ nor $\upsilon$ uniquely govern the dynamics: For the former (activated dynamics) the isothermal data would be horizontal lines, while for the latter (jamming dynamics) all data would superimpose to a single curve. Nevertheless, it is evident that for PC the effect of $\upsilon$ is weaker than for PCB54; that is, much larger changes in $\upsilon$ are necessary to obtain a given change in $\tau$.

With both $T$ and $\upsilon$ influencing the dynamics, we quantify their roles by using a scaling function recently shown to be valid for many glass-formers[21,22]

$$\log(\tau) = \Im(T\upsilon^{\gamma}) \qquad (2)$$

where $\gamma$ is a material specific constant. This relation, which has been verified by other groups experimentally[23,24] and by simulation[25], is a generalization of $\gamma=4$ as originally found for ortho-terphenyl[26,27]. It is also consistent with an analysis of NMR results for polymers.[28] When relaxation times measured at different $\upsilon$ and $T$ are plotted versus $T\upsilon^{\gamma}$, all data superimpose, as illustrated for PCB54 and PC in figure 1(b). The simplest interpretation of this behavior is to consider the intermolecular potential as the sum of a repulsive inverse power potential (with exponent $3\gamma$) and an attractive mean field[29]. While this interpretation may be not apply when strong attractive interactions are present, such as in hydrogen bonded materials, or for polymers, which have covalent bonds between segments, it does offer a starting point for linking molecular motions to an effective intermolecular potential. The parameter $\gamma$ can be regarded as a measure of the steepness of the potential.

Equation (2) also facilitates extension of the analysis of the dynamics to arbitrary thermodynamic condition, because once the EOS and $\gamma$ are known, $\tau$ is readily determined for any $T$ and $\upsilon$.[30] For example, in figure 2 the behavior at constant volume $\upsilon(T_g,P_{atm})=\upsilon_g$ is obtained by calculating for each value of $\log(\tau)$ the T conforming to the condition $T\upsilon^{\gamma}(0.1MPa) = T\upsilon_g^{\gamma}$ for PCB54 and PC.

The magnitude of the parameter $\gamma$ must reflect the relative contribution of $T$ and $\upsilon$ to the dynamics, $\gamma \to 0$ for thermally-activated motions and $\gamma$ large for jammed (or congested) dynamics. In fact $\gamma$ is related to another quantity commonly used for this purpose, the ratio of the activation enthalpy at constant $\upsilon$ ($E_V = \left.\frac{\partial \log(\tau)}{\partial(1/T)}\right|_V$) to the enthalpy at constant P ($E_P = \left.\frac{\partial \log(\tau)}{\partial(1/T)}\right|_P$)[21]

$$\left.\frac{E_V}{E_P}\right|_{T_g} = (1+\gamma\alpha T_g)^{-1} \quad (3)$$

where $\alpha$ is the isobaric thermal expansion coefficient at $T_g$. Defining $m_P$ as the isobaric fragility and $m_V$ as the isochoric (constant $\upsilon$) fragility, then from equation (1)

$$\frac{m_V}{m_P} = \frac{E_V}{E_P} \quad (4)$$

It follows that if $m_V = m_P$, then T is dominant, while $m_V \to 0$ when $\upsilon$ dominates.

In figure 3 we report all available $m_V$ and $m_P$ data, collected from various publications, and including results for molecular liquids, polymers, and hydrogen bonded glass-formers. From this figure it is evident that a strong correlation exists between the value of $m_V$ and $m_P$; we find by linear regression

$$m_P = (37 \pm 3) + (0.84 \pm 0.05) m_V \quad (5)$$

with a correlation coefficient = 0.95. Included in figure 3 are the lines for $m_V = m_P$ and $m_V = 0$, corresponding respectively to activated and jammed dynamics. All real materials fall between these two extremes. Since the magnitude of $m_V$ (or $m_P$) is directly related to $m_V/m_P$, we can calculate from eq (5) the limiting values of $m_P$: $m_P = 37 \pm 3$ for $m_V = 0$ and $m_P = 231 \pm 72$ for $m_V = m_P$. These correspond well to the range found experimentally at atmospheric pressure; for example, according to Böhmer et al.[4], $40 \leq m_P \leq 191$ for small molecules and polymers.

This analysis shows that the dynamics in fragile liquids is for the most part thermally activated, while congested dynamics predominates for strong liquids. Of course, this is only a general pattern, rather than a strict relationship, since details of the molecular structure may have secondary effects. For example, for the strongly associated glycerol, the $m_V/m_P$ ratio is large (= 0.94[31,32]) but the fragility is small ($m_P = 54$[32]). We expect hydrogen bonded materials as a class to exhibit deviations from the correlation between isochoric and isobaric fragilities. Likewise, the small fragilities observed for network glasses and orientationally disordered crystals ($m_P$ as low as 14) are not necessarily consistent with the correlation in figure 3, although no data are available to asses this. Therefore, presently the results herein are considered valid primarily for molecular and polymeric glass-formers, although inclusion of the two H-bonded liquids in figure 3 would not change the quality of the linear fit to the data.

As discussed above, $\gamma$ is a measure of the relative contribution of $T$ and $\upsilon$, which means that $\gamma$ should also be related to $m_V$. In figure 4, we have plotted $\gamma$ versus the inverse isochoric fragility for 26 materials, demonstrating the relatively strong (inverse) correlation between the two quantities - large $\gamma$ (strong effect of $\upsilon$) corresponding to small fragility and *vice versa*. The approximately linear behavior in fig. 4 follows from equations (3) to (5), together with the empirical rule of Boyer and Bondi[33] that the product of $\alpha T_g$ is approximately constant, $\cong 0.16$-$0.19$.

Since (as discussed above) the parameter $\gamma$ can be linked to the exponent of the intermolecular potential, the results in figures 3 and 4 suggest that the fragility has a similar origin. For a given material, a dominant short range repulsive potential gives rise to stronger (less fragile) dynamics. Larger $\gamma$ implies steeper potential wells (as depicted in fig.4 with the sketch taken from Angell[2]), and hence a liquid structure more resistant to changes in $T$. Relaxation is facilitated by changes of the energy barriers (from changes in intermolecular distances); thus, the effect of $\upsilon$ becomes more important for strong liquids. For fragile liquids, the potential energy surface is characterized by flatter minima (illustrated in fig.4), so that thermally activated motion can proceed. Evidently the shape of the potential affects its anharmonicity, a steeper potential (larger $\gamma$) being more harmonic. According to this interpretation, the fragility of liquids increases with the anharmonicity of the potential, an idea consistent with other results.[8,34]

In contrast, a simulation by De Michele et al.[35] found no effect of the strength of the intermolecular repulsive potential (i.e., γ) on the fragility. However, these simulations were for temperatures above the mode coupling critical temperature, and thus not directly relevant to the dynamics near $T_g$ of interest herein. Of course, our observed correlation between fragility and $\gamma$ is an experimental fact, notwithstanding any connection of the latter to the intermolecular potential. Inferring relationships between the supercooled dynamics and the topology of the intermolecular potential is the focus of many investigations into the glass transition.[1,2,26,36,37,38,39,40,41,42]

In conclusion, extensive experimental evidence is presented showing a linear correlation between the isobaric and isochoric fragility. This implies that the fragility of glass-formers is directly related to the relative contribution of $T$ and $\upsilon$ to the dynamics. A large fragility reflects the dominance of thermally activated dynamics, while for strong liquids, the dynamics is governed more by jamming (excluded volume among neighboring molecules or chain segments). These ideas are consistent with the scaling $\tau = \Im(T\upsilon^{\gamma})$, suggesting a connection between fragility and the steepness of the intermolecular potential, and consequently its anharmonicity.

**Acknowledgment**

This work was supported by the Office of Naval Research. We thank K.L. Ngai for stimulating discussions.

**Figure captions**

**Figure 1.** (a) Dielectric relaxation times for PCB54[19] and PC[20] as a function of specific volume at constant (atmospheric) pressure (solid symbols) and at constant temperature (open symbols). (b) Same data plotted vs the function $T^{-1}\upsilon^{-\gamma}$.

**Figure 2.** Dielectric relaxation times for PCB54 and PC at atmospheric pressure (solid symbols) and at constant $\upsilon = \upsilon_g$ (open symbols). Solid lines are the data at $\upsilon_g$ calculated from the atmospheric pressure data using the scaling relation $\tau = \Im(T\upsilon^\gamma)$.

**Figure 3.** Isobaric fragility $m_P$ (at atmospheric pressure) vs isochoric fragility $m_V$ for 33 materials (in order of increasing $m_V$): PCB62[30], 1,1'-di(4-methoxy-5-methylphenyl)cyclohexane (BMMPC)[30], 1,1'-bis(p-methoxyphenyl)cyclohexane (BMPC)[43], PCB54[19], PCB42[19], cresolphthalein-dimethylether (KDE)[30], salol[30], glycerol[31,32], phenylphthalein-dimethylether (PDE)[30], polypropylene oxide (PPO)[44], polymethylphenylsiloxane (PMPS)[45], o-terphenyl (OTP)[24], polyepichlorhydrin (PECh)[24], polymethyltolylsiloxane (PMTS)[46], polyvinylmethylether (PVME)[24], polyvinylacetate (PVAc)[44], polystyrene (PS)[47], polypropylene glycol (PPG)[48], PC[30], diglycidyl ether of bisphenol A (DGEBA)[44], 1,4-polyisoprene(PI)[49,50], poly[(phenol glycidyl ether)-co-formaldehyde] (PPGE)[44], PVAc(2)[24], polyvinylethylene (PVE)[51], 1,4-polybutadiene (PB)[24], polyethylacrylate (PEA)[47], polymethylacrylate (PMA)[44], PMA(2)[47], sorbitol[21,52], and polyvinylchloride (PVC)[47]. Where $m_V$ was not given, it was calculated using eq.(4). The lower left and upper right correspond to the respect extremes for $m_V$ and $m_P$. The solid line is the linear fit (correlation coefficient = 0.95).

**Figure 4.** Parameter $\gamma$ vs the inverse isochoric fragility for 26 materials. The data (in order of increasing $m_V$) are: PCB62, BMMPC, BMPC, PCB54, PCB42, KDE, salol, glycerol, PDE, PMPS, OTP, PECH, PMTS, PVME, PVAc, PPG, PC, DGEBA, PI, PPGE, PVAc(2), PVE, PB, PCGE, and sorbitol. Values of $\gamma$ are from references [19,20,21,22,23,24]. The solid line is a linear fit (correlation coef. = 0.92) to all data

except the H-bonded materials (if included correlation coef. = 0.88). Representations of the potential energy hypersurface taken from Angell[2]

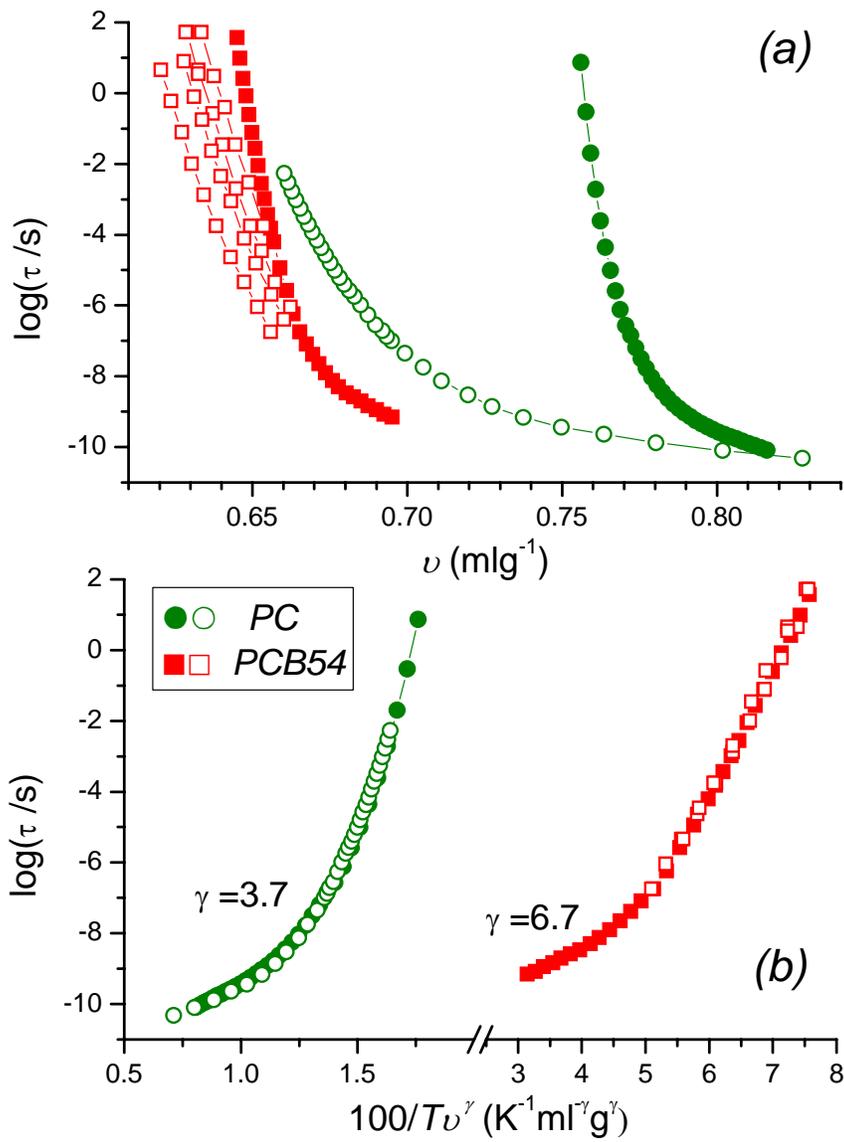

**Figure 1**

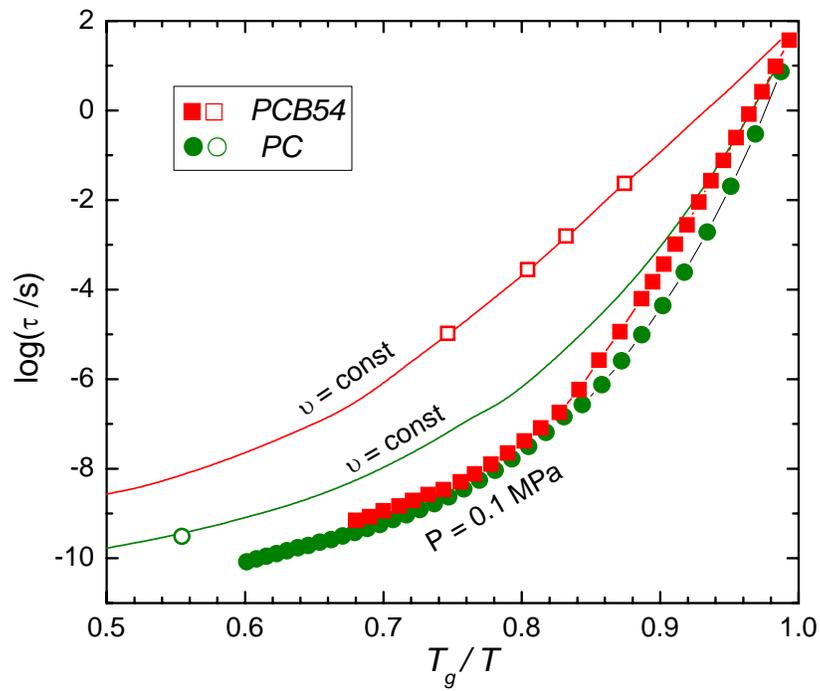

**Figure 2**

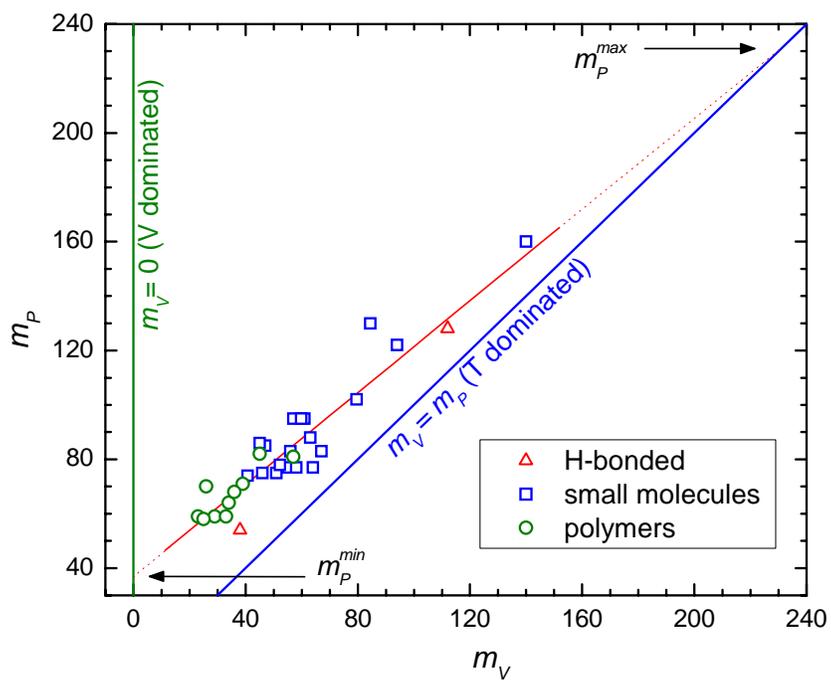

**Figure 3**

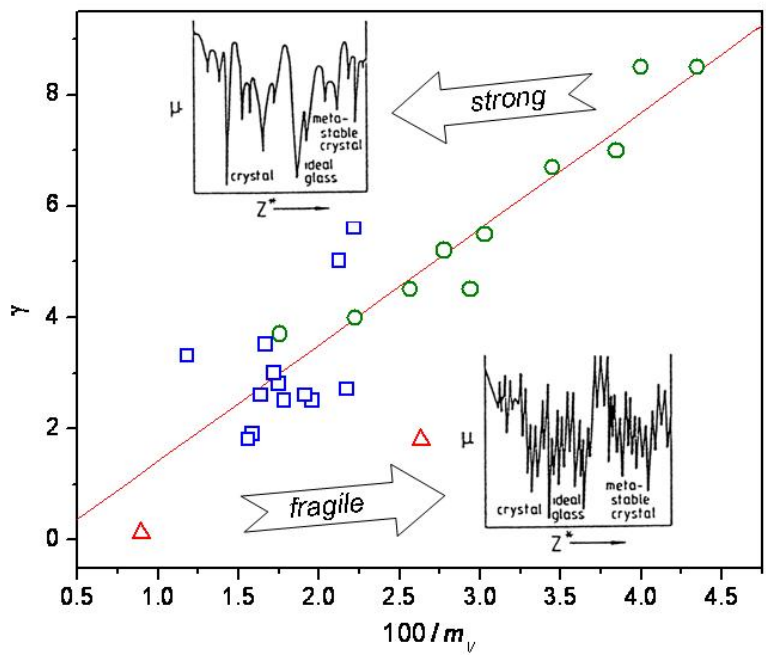

**Figure 4**